# Nonlinear domain wall velocity in ferroelectric Si-doped HfO$_2$ thin film capacitors

So Yeon Lim,[1] Min Sun Park,[2] Ahyoung Kim,[1] and Sang Mo Yang[1,a]

[1]*Department of Physics, Sogang University, Seoul 04107, Republic of Korea*

[2]*Department of Physics, Sookmyung Women's University, Seoul 04310, Republic of Korea*

[a]Author to whom correspondence should be addressed: smyang@sogang.ac.kr

**Abstract**

We investigate the nonlinear response of the domain wall velocity ($v$) to an external electric field ($E_{ext}$) in ferroelectric Si-doped $HfO_2$ thin film capacitors using piezoresponse force microscopy (PFM) and switching current measurements. We verified the reliability of the PFM images of ferroelectric domain switching by comparing the switched volume fraction in the PFM images with the time-dependent normalized switched polarization from the switching current data. Using consecutive time-dependent PFM images, we measured the velocity of the pure lateral domain wall motion at various $E_{ext}$. The $E_{ext}$-dependent $v$ values closely follow the nonlinear dynamic response of elastic objects in a disordered medium. The thermally activated creep and flow regimes were observed based on the magnitude of $E_{ext}$. With a dynamic exponent of $\mu = 1$, our thin film was found to have random-field defects, which is consistent with the Lorentzian distribution of characteristic switching time that was indicated in the switching current data.

Since the discovery of unexpected ferroelectricity in Si-doped $HfO_2$ in 2011,[1] doped $HfO_2$ thin films have attracted much attention in the fields of ferroelectricity and non-volatile memories. Unlike conventional perovskite ferroelectrics, fluorite-structure doped $HfO_2$ has many advantages from the application point of view, such as good Si-compatibility, extremely thin physical thickness, and large bandgap.[2, 3] Thus, in the last decade, many studies have been conducted to promote the realization of $HfO_2$-based ferroelectric memory and logic devices.[2-13] However, despite the tremendous research, direct measurements of domain wall velocity in $HfO_2$ thin film capacitors have been rarely performed.[6] Because the velocity of lateral domain wall motion is a crucial factor in determining the operation speed of ferroelectric-based devices, it is highly required for understanding of ferroelectric domain wall dynamics in doped $HfO_2$ thin film capacitors, in particular, the domain wall velocity ($v$) response to an external electric field ($E_{ext}$).

Polarization switching in ferroelectric thin films is known to be achieved by several stages that include nucleation, forward growth, sideways growth, and coalescence of domains.[14, 15] Among these stages, sideways domain growth (i.e., lateral domain wall motion) has been regarded as a problem to the pinning-driven propagation of elastic objects in disordered media with quenched defects.[15-19] In such disordered systems, the velocity of the elastic objects has an intriguing nonlinear dynamic response to external forces due to the competition between their elastic energy and the pinning potential produced by defects (see Fig. 1(b) of Ref. 16). In ferroelectric materials, different $v$ regimes exist depending on the magnitude of $E_{ext}$ at a finite temperature, such as the creep regime ($v \propto \exp[-(E_a/E_{ext})]^\mu$) under low $E_{ext}$, where $E_a$ and $\mu$ are the activation field and dynamic exponent, respectively, and the flow regime ($v \propto E_{ext}$) under high $E_{ext}$.[16] The thermally activated creep motion of ferroelectric domain walls has been studied in depth in epitaxial $Pb(Zr,Ti)O_3$ thin films using piezoresponse force microscopy (PFM).[18-20] Furthermore, Jo *et al*. first demonstrated that whole nonlinear domain wall dynamics (including creep, flow, and depinning regimes) are valid at wide temperature ranges in epitaxial Pb(Zr,Ti)O$_3$ thin films using switching current measurements combined with PFM imaging.[16] However, the $E_{ext}$-dependent nonlinear responses of ferroelectric domain wall dynamics has yet to be investigated in doped HfO$_2$ thin films despite its importance both fundamentally and technologically.

In this study, we investigate the nonlinear domain wall dynamics in ferroelectric Si-doped HfO$_2$ thin film capacitors using PFM combined with switching current measurements. We visualized ferroelectric domain nucleation and growth on top of the top electrodes using PFM with an additional probe needle (APN-PFM).[21] From the successive time-dependent PFM images, we measured the velocity of pure lateral domain wall motion at various $E_{ext}$. The $E_{ext}$-dependent $v$ values closely follow the nonlinear dynamic response, including the creep and flow motions. We experimentally determined the dynamic exponent to be $\mu = 1$, which indicates the random-field nature of the defects in our film.

As a model system for this study, we used a 4.2 mol% Si-doped HfO$_2$ thin film sandwiched between TiN electrodes (i.e., Pt/TiN/Si:HfO$_2$/TiN capacitors) and grown on a Si substrate. The 8-nm-thick Si:HfO$_2$ layer was fabricated by atomic layer deposition. The 10-nm-thick top and bottom TiN electrodes were grown by a chemical vapor deposition process. The thickness of the top Pt layer was 2 nm. The diameter of the patterned Pt/TiN top electrodes was 100 μm. Details of a sample fabrication can be found in existing literature.[7, 8] After the deposition of the top electrode, post-annealing was performed out at 900 °C for 1 s under ambient N$_2$ conditions. Polarization-electrical voltage (*P-V*) hysteresis loops were measured via a TF analyzer 2000E (aixACCT). For switching current measurements, a digital oscilloscope (DLM2024, Yokogawa) and arbitrary wave generators (FG400, Yokogawa) were used and controlled by a home-built software written in LabVIEW. For the APN-PFM, we attached a probe needle to a commercial atomic force microscope (NX10, Park Systems), for which a similar setup is reported in other publications.[20, 22] We used non-conductive cantilevers (PPP-FMR, Nanosensors) for the PFM imaging, minimizing artifacts such as electrostatic interaction.[23, 24] For the PFM measurements, an AC bias of 0.4 V with near contact resonance frequency (~ 380 kHz) was applied to the top electrode via the attached probe needle. We obtained PFM images with a high signal-to-noise ratio using the dual-amplitude resonance tracking (DART)[25] technique with a lock-in amplifier (HF2LI, Zurich Instrument).

We first measured the basic ferroelectric properties of the Si:HfO$_2$ capacitors, such as the remnant polarization ($P_r$) and coercive voltage ($V_c$). Figures 1(a) and 1(b) show the $P$-$V$ hysteresis loops and the corresponding current-voltage ($I$-$V$) curves of the Si:HfO$_2$ capacitors, respectively. All data were measured by triangular pulses with a frequency of 1 kHz and an amplitude of ±2.5 V. To check the effects of electric field cycling on the ferroelectric properties, i.e., the so-called wake-up effect,[9, 10, 26] 1000 cycles of rectangular pulses at an amplitude of ±2.5 V and a frequency of 20 kHz were applied to the pristine specimen. Note that the $P$-$V$ and $I$-$V$ data did not show any further change after 1000 cycles (up to ~ 10$^5$ cycles, not shown). As shown in Fig. 1(a), the wake-up effect was not significant in our Si-doped HfO$_2$ capacitors in terms of the $P_r$ value. This is presumably due to the relatively high annealing temperature. In both the pristine and the woken-up states, the 2$P_r$ values were approximately 36 $\mu$C/cm$^2$. However, $V_c$ showed a slight change after the wake-up process. Specifically, in the pristine state, the $V_{c+}$ and $V_{c-}$ values were -1.21 V and 0.45 V, respectively. This negative imprint behavior became weak in the woken-up state with $V_{c+}$ and $V_{c-}$ values at -1.09 V and 0.56 V, respectively. This horizontal imprint can be ascribed to the difference in concentrations of oxygen vacancies and/or trap sites at the top and bottom interfaces.[7, 12] However, the $I$-$V$ curves showed more complicated changes such as the shift in the peak positions and the enhancement in the peak sharpness, as shown in Fig. 1(b). This change in the $I$-$V$ curves implies that the ferroelectric domain switching dynamics can vary in the early switching stage (i.e., during the 1000 cycles), although the overall polarization value did not significantly change. Note that we thus performed switching current measurements and PFM imaging only on the woken-up capacitors.

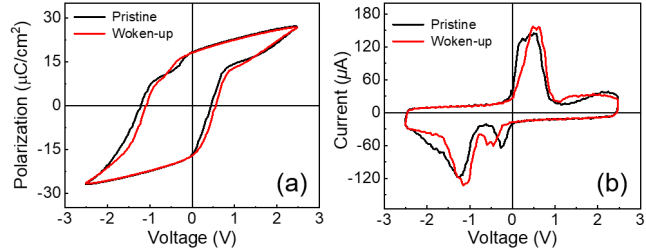

**FIG. 1.** (a) *P-V* hysteresis loops and (b) corresponding *I-V* curves for the Pt/TiN/Si:HfO$_2$/TiN capacitors in the pristine state (black) and the woken-up state (red).

Before conducting the spatially-resolved PFM imaging of the nanoscale ferroelectric domains, we performed transient switching current measurements to obtain the macroscopic information regarding the polarization switching kinetics. The detailed procedure for the switching current measurements that we performed can be found in other literature.[21] Note that in this study, the rectangular pulses with a ±2.5 V amplitude and a 1 ms duration were used as the poling and reading pulses. The other conditions were equivalent to those in Ref. 21. Figures 2(a) and 2(b) show the time (*t*)-dependent normalized switched polarization ($\Delta P(t)/2P_r$) at different external voltages ($V_{ext}$) for the positive and negative biases, respectively. Here, *t* indicates the duration of the writing pulse for polarization partial switching. The polarization switching occurred earlier in time and $\Delta P(t)/2P_r$ approached larger values with an increase in the magnitude of $V_{ext}$ for both biases. We found that the $V_c$ value is an important parameter for determining the speed of the overall polarization switching. For the positive and negative biases, we can divide the two groups categorized by the data at $V_{ext}$ = 0.5 V (~ $V_{c+}$) and at $V_{ext}$ = -1.0 V (~ $V_{c-}$), respectively. In other words, when $|V_{ext}| < |V_c|$, the speed of macroscopic polarization switching is significantly reduced.

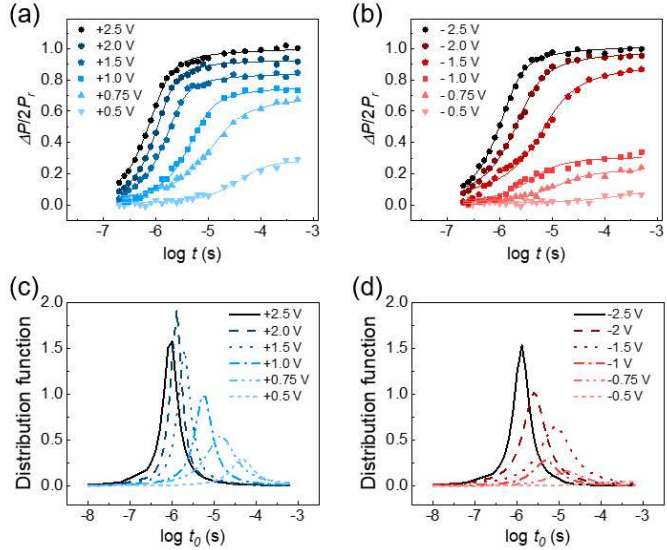

**FIG. 2.** Time-dependent $\Delta P(t)/2P_r$ data as a function of different $V_{ext}$ for the (a) positive and (b) negative biases. The solid lines in (a) and (b) are the fitting results based on the NLS model with a Lorentzian distribution of log $t_0$. The $V_{ext}$-dependent Lorentzian distribution functions for the (c) positive and (d) negative biases.

For further insight into the polarization switching kinetics, we applied the nucleation-limited switching (NLS) model,[27] which has been widely used to explain polarization switching in polycrystalline ferroelectric films. In the NLS model, $\Delta P(t)$ is described as:

$$\Delta P(t) = 2P_r \int_{-\infty}^{\infty} [1 - \exp\{-(t/t_0)^n\}] F(\log t_0) d(\log t_0), \qquad (1)$$

where $t_0$ and $n$ are the characteristic switching time and effective dimension for domain growth, respectively, and $F(\log t_0)$ is the distribution function of the logarithmic characteristic switching time ($t_0$).[27] This model indicates that the polycrystalline ferroelectric films can be regarded as an aggregation consisting of many switching regions with different $t_0$. Consequently, understanding the distribution function of $t_0$ is crucial.

We found that the polarization switching kinetics in our Si:HfO$_2$ capacitors can be well explained by the NLS model. The solid lines in Figs. 2(a) and 2(b) are the fitting results based on the Lorentzian distribution of log $t_0$:

$$F(\log t_0) = \frac{A}{\pi}\left[\frac{w}{(\log t_0 - \log t_1)^2 + w^2}\right], \quad (2)$$

where $A$ is the normalization constant, $w$ is the half-width at half-maximum, and log $t_1$ is the central value of the distribution function.[28] Note that the fitting results agree well with the experimental $\Delta P(t)/2P_r$ data. Overall, log $t_1$ and $w$ decrease with increasing $V_{ext}$ for both biases, as displayed in Figs. 2(c) and 2(d). According to Jo *et al.*,[28] the Lorentzian distribution of log $t_0$ is due to the local field variation originating from defect dipoles at pinning sites. Therefore, it can be concluded that the Si-doped HfO$_2$ capacitors have the variations of the local field, resulting in a complicated energy landscape for domain wall motion. Such an energy landscape will induce pinning-driven nonlinear domain wall dynamics.

To directly investigate ferroelectric domain wall dynamics at the nanoscale, we performed PFM imaging in the capacitor geometry using APN-PFM.[21] With the rectangular pulse trains of the stroboscopic PFM method,[29] we measured the successive PFM images on the $t$-dependent ferroelectric domain switching, as shown in Figs. 3(a) – 3(c). Specifically, we first applied the reset pulse (±2.5 V, 1 ms) to set the initially poled state to a single polarization. Then, we applied the writing pulse for polarization partial switching (with the opposite polarity as the reset pulse) and measured the PFM images. These procedures were repeated with an increase in the duration of the writing pulse.

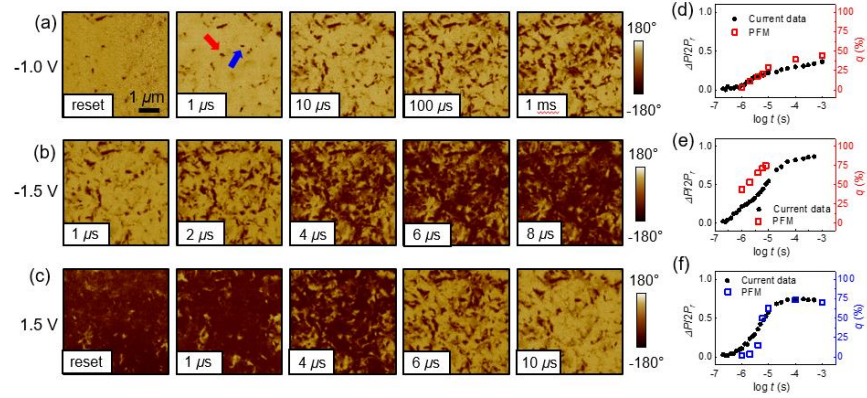

**FIG. 3.** Successive PFM phase images of ferroelectric domain evolution at (a) $V_{ext}$ = -1.0 V, (b) -1.5 V, and (c) 1.5 V. All images were measured at the same sample area. The scan size was 5 × 5 μm². Time-dependent $\Delta P/2P_r$ (solid circles) and $q$ (open squares) values measured at (d) $V_{ext}$ = -1.0 V, (e) -1.5 V, and (f) 1.5 V.

By comparing the *t*-dependent PFM images, we obtained spatially-resolved information of nucleation and domain wall motion as well as pinned domains at the nanometer level. As shown in the reset images of Figs. 3(a) and 3(c), most of the regions were switched after the application of the reset pulse. However, unswitched small domains can still be observed. This suggests that even in the woken-up capacitors, some domain walls are strongly pinned by defects (such as defect dipoles and/or structural grain boundaries) and thus they cannot be fully switched. By increasing the duration of the writing pulses, it can be observed that multiple domains nucleated, subsequently grew, and finally merged together. Although many domains were newly created (e.g., blue arrow in Fig. 3(a)), some domains (red arrow) seemed to grow from the residual unswitched (pinned) domains. Similar behavior was reported in the PFM study of La-doped HfO₂ capacitors.[6] At $V_{ext}$ = -1.0 V ($|V_{ext}| < |V_{c-}|$), the domain walls propagated slowly. Even after applying the writing pulse for 1 ms, the switched region encompassed less than 50%. However, at $V_{ext}$ = -1.5 V ($|V_{ext}| > |V_{c-}|$), the domains grew faster. After only 8 μs, most of the regions were switched. This trend in domain switching agrees well with that of switching current data. The reliability of the measured PFM images was confirmed by comparing the switched volume fraction $q$ from the PFM images (open squares) with $\Delta P/2P_r$ from the switching current data (solid circles), as shown in Figs. 3(d) – 3(f). The $q$ values were estimated by integrating $R\cos\theta$, where $R$ and $\theta$ are the values of the PFM amplitude and phase, respectively.[21] Despite the slight discrepancy, the two data matched well, demonstrating that the local PFM images well represent the domain switching behavior of the Si:HfO$_2$ capacitors.

Intriguingly, although the polarity and duration of the applied bias were different, we observed similar PFM images (in terms of switched domain patterns) in the intermediate stage of domain switching. For example, the PFM image at $V_{ext}$ = -1.0 V and $t$ = 100 μs is almost identical to that at $V_{ext}$ = -1.5 V and $t$ = 2 μs. Also, it is similar to the image at $V_{ext}$ = 1.5 V and $t$ = 6 μs. It is known that in ferroelectric systems, nucleation occurs at particular sites presumably due to defects (the so-called inhomogeneous nucleation).[22] Thus, for the cases under the same polarity bias, inhomogeneous nucleation can be a major reason for the similarity between the PFM images. However, the opposite polarity cases cannot be simply explained by inhomogeneous nucleation. We suggest that this intriguing behavior originates from the occurrence of nucleation-limited switching. If the film is composed of many small switching regions and $t_0$ is predetermined for each region and bias polarity, the stochastic nature of the domain wall motion becomes significantly reduced and, consequently, similar PFM images can be observed in the intermediate stage regardless of the polarity and duration of the applied switching bias.

Finally, we estimated the $t$-dependent $v$ values from the successive PFM images. The inset of Fig. 4(a) shows the $t$-dependent evolution of a nucleated domain under a bias of -1.5 V. To measure the $v$ values, for simplicity, we approximated that the domain had an elliptical shape. From the $t$-dependent change in the median values of the major and minor axes, we calculated the $v$ values. The estimation of $v$ stopped when a domain merged with another domain.

Therefore, our approximation is applicable to a relatively small domain in the early stages of switching process. The detailed procedure for the $v$ calculation is described in the supplementary material. Because we calculated the growing area from the nucleated domain, we were able to obtain the velocity of the pure lateral domain wall motion. Figure 4(a) shows $t$-dependent $v$ values at $V_{ext}$ = -1.5 V. The $v$ value decreased with increasing $t$. The maximum velocity was approximately 0.10 m/s, which is similar to that in the La:HfO$_2$ capacitors.[6] There was a significant drop in $v$ when the domain reached a certain size (~ a few hundred nm), presumably due to the strong pinning potential. Compared with the average grain size (~ 50 nm), it can be concluded that the domain walls can propagate across several grain boundaries.[6] It should be noted that the elliptical shape approximation does not reflect the fractal nature of domain walls, which is an important characteristic of ferroelectrics.[30-33] The domain movement is non-smooth due to the pinning-depinning transitions (i.e., sudden jumps called jerks) of domain walls, leading to fractal domain patterns.[30-33] Note that, in this study, we focused on the acquisition of average domain wall velocity from multiple domains using a simple approximation.

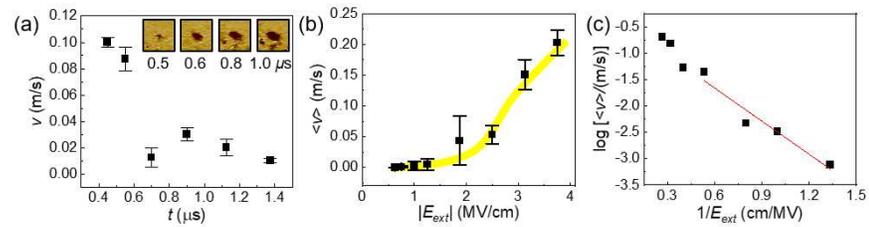

**FIG. 4.** (a) The <$v$> values as a function of $t$ under a bias of -1.5 V. The inset shows the successive enlarged snapshots of a growing domain. The image size is 0.3 × 0.3 μm$^2$. (b) The $E_{ext}$-dependent <$v$> values from the PFM images measured at the negative bias. The solid (yellow) line is only for guideline of eye. (c) The <$v$> values as a function of the inverse $E_{ext}$. The solid (red) line is the linear fitting result.

The $E_{ext}$-dependent average lateral domain wall velocity (<$v$>) was strongly nonlinear, as shown in Fig. 4(b). The plot of <$v$>-$E_{ext}$ resembles Fig. 1(b) of Ref. 16. This indicates that the lateral domain wall motion in the polycrystalline Si:HfO$_2$ capacitors closely follows the nonlinear dynamic responses, which are governed by the competition between the pinning potential and the elastic energy of domain walls. Under a low $E_{ext}$ regime (i.e., less than the coercive field), $v$ appears to follow a thermally activated creep motion. Under a high $E_{ext}$ regime (the two highest $E_{ext}$ values: |$V_{ext}$| = 2.5 V and 3.0 V), $v$ corresponds to the flow regime. Due to the thermal effects of room temperature, the smooth transition from the creep regime to the flow regime is observed in the intermediate $E_{ext}$ regime.

To further investigate the creep motion of the domain walls, a linear fitting was applied to the plot of log <$v$> and 1/$E_{ext}$ in the low $E_{ext}$ regime, as displayed in the solid (red) line of Fig. 4(c). The fitting results agree well with the experimental data, producing the values of $\mu$ = 1 and $E_a$ ~ 4.81 MV/cm. According to statistical physics, the dynamic exponent $\mu$ can indicate the nature of the pinning potential by defects such as random-field or random-bond.[18] The value of $\mu$ = 1 indicates that our film shows random-field disorder, which is a long-range pinning potential due to defects.[15, 18] This result is also consistent with the fact that the Lorentzian distribution of log $t_0$ implies $\mu$ = 1.[28] Note that the estimated $E_a$ value of 4.81 MV/cm is similar to the $E_a$ values measured from macroscopic switching current data in (Hf,Zr)O$_2$ thin films.[13] However, it is 6 – 7 times larger than that of epitaxial Pb(Zr,Ti)O$_3$ capacitors.[20] The HfO$_2$-based thin films have a significantly higher $E_a$ value and consequently a smaller $v$ value compared to perovskite ferroelectrics, which is consistent with a recent theoretical calculation on the flat polar phonon bands in HfO$_2$.[34]

In summary, we investigated the nonlinear domain wall velocity in ferroelectric Si:HfO$_2$ thin film capacitors using PFM and switching current measurements. We measured the $v$ values of pure lateral domain wall motion at various $E_{ext}$ from $t$-dependent PFM images. The $E_{ext}$- dependent $v$ values closely followed the nonlinear domain wall dynamics, clearly showing the creep and flow regimes. We found that our thin film contained random-field defects, as indicated by the dynamic exponent $\mu = 1$. These PFM results were consistent with the switching current data described by the NLS model with the Lorentzian distribution of log $t_0$. The direct measurement of $v$ in the nanoscale domain evolution images gives insight into the nonlinear dynamics of domain walls and the nature of pinning potentials as well as the optimization of the operating speed in ferroelectric $HfO_2$-based devices.

**Supplementary Material**

See supplementary material for the procedure for the calculation of domain wall velocity.

**Acknowledgements**

The authors gratefully acknowledge H. K. Yoo and D. I. Suh for the sample. This work was supported by the National Research Foundation of Korea (NRF) grant funded by the Korean Government (MSIP) (No. NRF-2019R1A2C1085812 & NRF-2017R1C1B2010258). This work was also supported by MOTIE (Ministry of Trade, Industry &Energy) (no. 10080657) and KRSC (Korea Semiconductor Research Consortium) support program for the development of future semi-conductor devices.

**Data availability**

The data that support the findings of this study are available from the corresponding author upon reasonable request.

Applied Physics Letters

ACCEPTED MANUSCRIPT

This is the author's peer reviewed, accepted manuscript. However, the online version of record will be different from this version once it has been copyedited and typeset.

PLEASE CITE THIS ARTICLE AS DOI: 10.1063/5.0035753

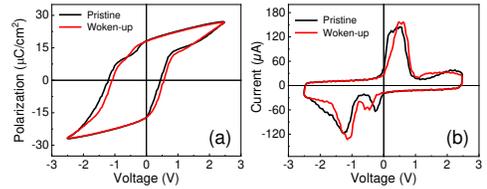

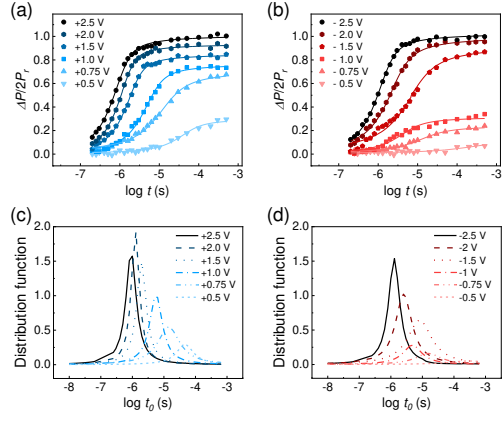

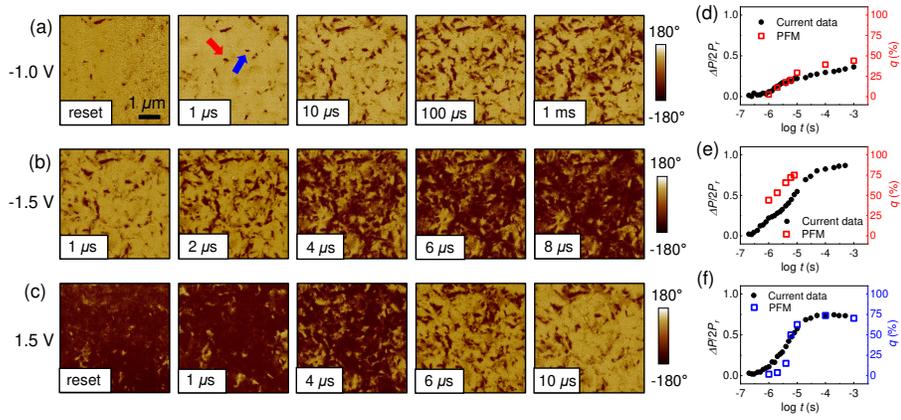

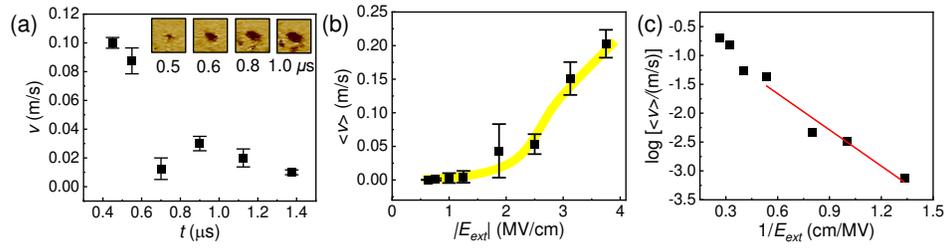